\documentclass[pra,twocolumn,floatfix,showpacs]{revtex4}
\usepackage{graphicx}

\newlength{\onecolfig}
\newlength{\twocolfig}
\newcommand{\ion}[2]{\mbox{$^{#2}$#1$^+$}}
\newcommand{\Ca}[1]{\ion{Ca}{#1}}

\newcommand{\lev}[2]{\mbox{#1$_{\mbox{\tiny$#2$}}$}}

\newcommand{\unit}[1]{\,\mbox{#1}}

\newcommand{\um}{\unit{$\mu$m}}
\newcommand{\nm}{\unit{nm}}

\newcommand{\ms}{\unit{ms}}
\newcommand{\us}{\unit{$\mu$s}}

\newcommand{\etal}{{\em et al.}}

\newcommand{\ish}{\mbox{$\sim$}\,}
\newcommand{\ltish}{\protect\raisebox{-0.4ex}{$\,\stackrel{<}{\scriptstyle\sim}\,$}}
\newcommand{\gtish}{\protect\raisebox{-0.4ex}{$\,\stackrel{>}{\scriptstyle\sim}\,$}}

\begin{document}
\bibliographystyle{apsrev}

\title{Scalable simultaneous multi-qubit readout with 99.99\% single-shot fidelity}

\author{A.~H.~Burrell, D.~J.~Szwer, S.~C.~Webster and D.~M.~Lucas}
\affiliation{Department of Physics, University of Oxford, Clarendon Laboratory, Parks Road, Oxford OX1 3PU, U.K.}

\date{submitted 23 June 2009; this version 12 March 2010}

\begin{abstract}
We describe single-shot readout of a trapped-ion multi-qubit register using space and time-resolved camera detection. For a single qubit we measure 
$0.9(3)\times 10^{-4}$ readout error in 400\us\ exposure time, limited by the qubit's decay lifetime. For a four-qubit register (a ``qunybble'') we 
measure an additional error of only $0.1(1)\times 10^{-4}$ per qubit, despite the presence of 4\% optical cross-talk between neighbouring qubits. A 
study of the cross-talk indicates that the method would scale with negligible loss of fidelity to \ish 10000 qubits at a density \ltish 1 
qubit/\um$^2$, with a readout time \ish 1\us/qubit.
\end{abstract}

\pacs{03.67.-a, 37.10.Ty, 42.50.Dv, 42.30.Va}

\maketitle

Single-shot qubit readout is an essential requirement for quantum computing, in particular for implementing quantum error-correction~\cite{07:Steane} 
or measurement-based quantum computing~\cite{06:Nielsen}. Fault-tolerant quantum computing requires readout fidelities \gtish 99.9\% if a large 
increase in the number of physical qubits per logical qubit is to be avoided~\cite{05:Knill}. Two major issues in scaling up from single-qubit to 
many-qubit systems are the need to perform operations on many qubits in parallel~\cite{07:Steane}, and to overcome errors due to ``cross-talk'' 
between different qubits. In this paper we describe simultaneous readout on a multi-qubit register and show that the quantum state can be accurately 
inferred even in the presence of significant cross-talk. Since the latter is short-range, the method is scalable to much larger numbers of qubits.

For any physical implementation, readout error is usually limited by the time available to integrate a detector signal (to improve the signal-to-noise 
ratio) before the qubit decays. In recent work on a two-electron spin qubit in a double quantum dot, for example, the trade-off was between electronic 
measurement noise and the 34\us\ qubit lifetime~\cite{09:Barthel}. Cross-talk in a multi-qubit system could increase the measurement noise and/or 
decrease the qubit lifetime. In trapped-ion or neutral-atom approaches to quantum computing, qubit readout is usually achieved by driving a 
fluorescing cycling transition which involves one of the qubit states but not the other, and measuring whether or not the atom 
fluoresces~\cite{80:Wineland}. Here the noise is predominantly photon-counting shot noise and the qubit lifetime is limited by spontaneous decay or 
off-resonant excitation, which can transfer the qubit from the ``dark'' state to the ``bright'' state or vice versa. Previous work reported 
measurement fidelities of 90\%--98\% for two- and three-qubit systems~\cite{06:Steffen,07:Hume,06:Acton}.

\begin{figure}[b]
\includegraphics[width=\onecolfig]{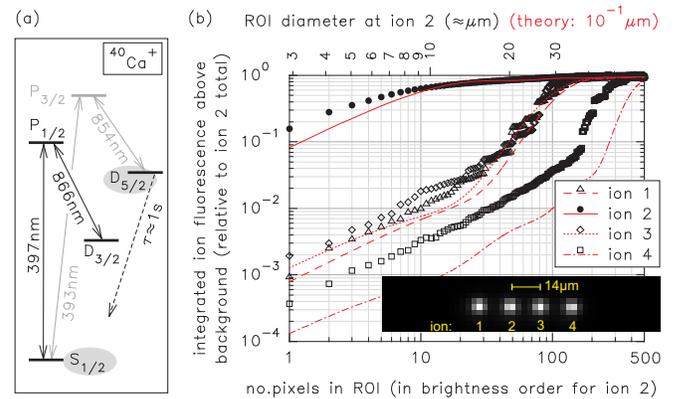}
\caption{%
(a) \Ca{40} levels. (b) Image cross-talk affecting ion 2 in a string of four ions (inset: $50\times 10$ pixel acquisition area, at scale 
2.6\um/pixel). Filled circles show the integrated fluorescence from ion 2; the decreasing-brightness pixel order and data were obtained by averaging 
images in which only ion 2 was fluorescing. For the other three ions (hollow symbols) the same pixels were taken in the same order, but the data were 
taken from images in which only that ion was fluorescing. Up to \ish 100 pixels, the region-of-interest (ROI) is roughly circular, centred on ion 2, 
with a diameter in the object plane given by the upper abscissa. The lines show calculated performance for a diffraction-limited imaging system of the 
same numerical aperture, with ion spacings {\em ten times smaller} than in the image, on a $10\times$ magnified upper abscissa.
}
\label{F:Xtalk}
\end{figure}

The highest fidelity quantum logic gate demonstrated to date~\cite{08:Benhelm} used a trapped-ion optical qubit stored in the (4\lev{S}{1/2}, 
3\lev{D}{5/2}) levels of \Ca{40}, fig.~\ref{F:Xtalk}a. For the same qubit, we have already described single-shot readout with 99.991(1)\% fidelity for 
a single ion, using time-resolved detection of photons counted with a photomultiplier tube (PMT)~\cite{08:Myerson}. Here, we extend this to 
simultaneous readout of multiple qubits using an electron-multiplying charge-coupled device (EM-CCD) camera, which allows both space- and 
time-resolved detection. We first measure the readout error for a single qubit and find that, as with the PMT, at any given photon collection 
efficiency this is limited by the decay lifetime $\tau=1168(7)\ms$ of the metastable \lev{D}{5/2} level~\cite{00:Barton}. Then we measure the readout 
error for a four-ion ``qunybble''~\cite{04:Larsson}, where we eliminate the effect of the \lev{D}{5/2} decay by post-selecting data, to find the 
residual error due to cross-talk between the imperfectly imaged ions. 

The linear ion trap apparatus~\cite{00:Barton} and the methods for preparing and detecting the bright \lev{S}{1/2} and dark \lev{D}{5/2} qubit states 
are similar to those in~\cite{08:Myerson}, with the PMT replaced by the camera. Fluorescence at 397\nm\ from the ion(s) is collected by a compound 
objective of numerical aperture $\sin\alpha=0.25$ and imaged onto the camera with net efficiency 1.0(1)\%. The objective is not optimized for 397\nm, 
resulting in performance well below the diffraction limit (fig.~\ref{F:Xtalk}b). The nominal quantum efficiency (QE) of the camera at 397\nm\ is 48\%, 
approximately twice that of the PMT. However, the stochastic gain process in the camera's EM register leads to an excess noise factor of $\sqrt{2}$ 
above shot noise when operated at high gain, equivalent to an effective reduction in QE by a factor 2 for photon counting 
applications~\cite{01:Mackay}. Background counts are dominated by the camera's clock-induced charge (CIC) readout noise rather than by scattered 
light, but the finite qubit lifetime is still the principal limitation to readout fidelity. Cosmic rays cause fewer errors for the CCD than the 
PMT~\cite{08:Myerson}, due to its spatial discrimination.

We define the average readout error per qubit as $\epsilon=\frac{1}{2}(\epsilon_B+\epsilon_D)$, where $\epsilon_B$ is the fraction of experiments in 
which an ion prepared in the bright state was detected to be dark, and similarly for $\epsilon_D$. In the single-qubit experiments described below, 
there is a small population preparation error for the dark state~\cite{08:Myerson}, resulting in a contribution to $\epsilon$ of $0.13(1)\times 
10^{-4}$; values of $\epsilon$ given below are after subtraction of this quantity. In the four-qubit experiment the state preparation error is 
expected to be negligible because post-selection is used.

\begin{figure}[t]
\includegraphics[width=\onecolfig]{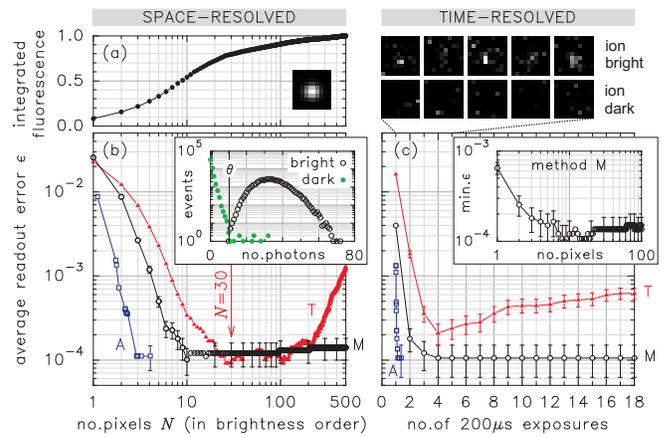} 
\caption{%
Single-ion readout results for space- and time-resolved detection. Analysis methods: (T)hreshold, (M)aximum likelihood, (A)daptive maximum likelihood. 
(a) Integrated fluorescence signal and long-exposure image. (b) Single 400\us\ exposure: readout error $\epsilon$ versus ROI size $N$ (inset: bright 
and dark photon count histograms for $N=30$, with threshold $\theta$). (c) Time-resolved detection using multiple 200\us\ exposures with, above, 
example exposures 1--5 for bright and dark state preparation (inset: minimum $\epsilon$ versus $N$ for method M).
}
\label{F:1ion}
\end{figure}

The readout fidelity for a single ion was measured by repeatedly preparing and measuring each qubit state, and comparing the measurement result with 
the known state preparation, for a total of 103744 trials. The method is similar to that described in~\cite{08:Myerson}, except that the time-resolved 
PMT detection bin is replaced by a single 400\us\ camera exposure. Camera images were analysed both by simple pixel-count thresholding and by maximum 
likelihood techniques. We label the pixels ${i=1\ldots N}$ in order of decreasing average brightness (determined from a long-exposure image, 
fig.~\ref{F:1ion}a) and select a region-of-interest (ROI) by varying $N$.

For the threshold analysis method, we histogram the total pixel counts $\sum_{i=1}^{N} n_i$ where $n_i$ is the photon count for pixel $i$, separately 
for the bright and dark state preparations, and choose the count threshold which optimizes discrimination between the histograms (fig.~\ref{F:1ion}b, 
inset). The variation of $\epsilon$ with ROI size $N$ is shown in fig.~\ref{F:1ion}b; it decreases with $N$ as the histogram overlap falls, reaches a 
minimum $\epsilon=0.9(3)\times 10^{-4}$ (with $\epsilon_B=0$) for $N=28$, then eventually rises when the ROI is too large due to increasing background 
counts from pixels far from the ion which contain negligible information.

The simple threshold method does not take into account the known spatial distribution of the ion fluorescence over the ROI. In the ``spatial'' maximum 
likelihood analysis, we consider the likelihood $p_B = \prod_{i=1}^N B_i(n_i)$ that the set of pixel counts $\{n_i\}$ came from a bright ion, and 
compare this to the likelihood $p_D = \prod_{i=1}^N D_i(n_i)$ that $\{n_i\}$ arose from a dark ion. Here $B_i(n_i)$ is the probability of observing 
$n_i$ counts on pixel $i$ for a bright ion; similarly for $D_i$. Since the probability distribution $B_i$ is different for each pixel, the information 
contained in the spatial distribution of the fluorescence is used. The distributions $B_i$ and $D_i$ can be obtained from independent control data, or 
by fitting the distributions from the experiment itself; either approach gave similar results. Maximum likelihood analysis results are also shown in 
fig.~\ref{F:1ion}b; the minimum value of $\epsilon=1.0(4)\times 10^{-4}$ is achieved with an ROI of only $N=10$ pixels, demonstrating that the maximum 
likelihood technique is more efficient than the threshold method. However, it does not lead to a lower $\epsilon$ as this is limited by the 
\lev{D}{5/2} decay.

An adaptive detection method (analogous to that described in~\cite{07:Hume,08:Myerson}) can reduce further the {\em average\/} number of pixels 
$\overline{N}$ required to determine the qubit state. The {\em estimated} probability of measurement error can be inferred from $p_B$ and $p_D$. For 
reliable readout, we only need use sufficient pixels for the absolute log-likelihood ratio $R = |\ln(p_B/p_D)|$ to be above some chosen confidence 
level. Taking the pixels in brightness order ensures most efficient use of the information. This method reaches the same $\epsilon$ in 
$\overline{N}=2.9$ pixels, fig.~\ref{F:1ion}b.

We also investigated time-resolved detection~\cite{Th:Langer} for a single qubit. In this experiment, the single 400\us\ camera exposure was replaced 
by a series of 18 exposures, each of duration $t_s=200\us$, to investigate whether time-resolved detection could give an improvement in readout 
fidelity by detecting any of the \lev{D}{5/2} decays which limited the single-exposure fidelity. The method is similar to that used for PMT detection 
in~\cite{08:Myerson}, except that the CIC noise associated with reading each exposure from the CCD means that it is disadvantageous to make each 
exposure too short. The CCD itself was used to store all 18 exposures, allowing a rapid sequence of images with minimal dead time (6\us) between each. 
Alternate bright and dark state preparations gave a total of 67892 trials ($\approx$ 1.2 million exposures), which were analysed by simple 
thresholding and by maximum likelihood analysis taking into account the effect of \lev{D}{5/2} decay, with results shown in fig.~\ref{F:1ion}c. For 
the ``spatio-temporal'' maximum likelihood analysis we compared the likelihoods $p_B = \prod_{j=1}^M p_{Bj}$ and
\[ 
p_D = \left(1-\frac{M t_s}{\tau}\right) \prod_{j=1}^M p_{Dj}
    + \left(\frac{t_s}{\tau}\right) \sum_{j'=1}^M \prod_{j=1}^{j'-1} p_{Dj} \prod_{j=j'}^M p_{Bj}  
\]
which allows for the probability $(t_s/\tau)$ of spontaneous decay in exposure $j'$ of $M$ (see~\cite{08:Myerson}). The probability $p_{Bj}$ that 
exposure $j$ came from a bright ion is given by $p_{Bj}=\prod_{i=1}^N B_{ji}(n_{ji})$ where $B_{ji}(n_{ji})$ gives the probability of measuring 
$n_{ji}$ counts on pixel $i$ of exposure $j$ for a bright ion; similarly for $p_{Dj}$. The minimum error is $\epsilon = 1.1(4)\times 10^{-4}$, for 
$M\geq 4$ exposures, no lower than the single-exposure experiment. A (temporally) adaptive analysis is also shown in fig.~\ref{F:1ion}c, and reaches 
the same $\epsilon$ in an average of $\overline{M}=1.1$ exposures (230\us); however the overhead in CCD read time for multiple exposures means the 
adaptive technique offers little practical speed-up.

We now describe an experiment to measure the readout error $\epsilon_X$ for the four-ion qunybble, independently of the \lev{D}{5/2} decay error. The 
cross-talk due to imperfect imaging in our system is illustrated in fig.~\ref{F:Xtalk}b, where the data were taken from the experiment to be described 
below. For an ROI centred on ion 2 with diameter equal to the mean ion spacing of 14\um, the nearest neighbour ions each contribute signals $\approx 
4.0\%$ of the signal from ion 2 itself, while the next-nearest neighbour contributes 0.9\%. Cross-talk at this level could clearly impact the readout 
error at the $10^{-4}$ level achieved for a single ion. 

We are not able to prepare all 16 possible qunybble states $\{0000,\ldots,1111\}$ deterministically with an error $\ish 10^{-5}$ as in the 
single-qubit experiments above. Instead, in each trial we prepare a random qunybble state, and sandwich the readout test measurement between ``pre'' 
and ``post'' measurements. We then select for analysis only those trials in which the ``pre'' and ``post'' measurements agree: these measurements give 
the state that was prepared, to which the ``test'' measurement is compared. A dark$\rightarrow$bright decay from the \lev{D}{5/2} level during the 
trial will cause the ``pre'' and ``post'' measurements to disagree for the ion which decayed. By increasing the exposure time of the ``pre'' and 
``post'' measurements only, we increase the accuracy of this state post-selection procedure at the expense of rejecting more trials.

\begin{figure}
\includegraphics[width=\onecolfig]{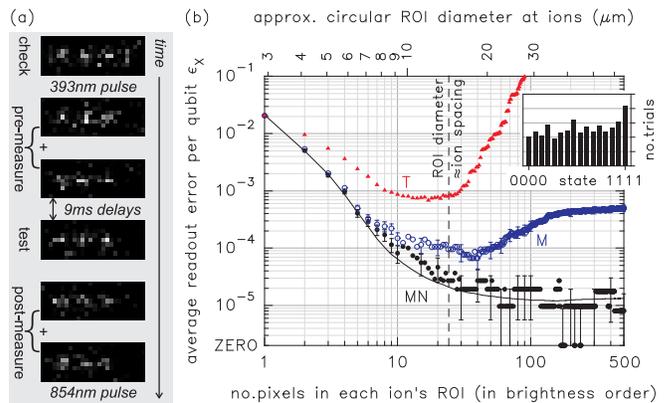}
\caption{%
Four-ion readout experiment. (a) Experimental sequence, showing a typical trial; maximum likelihood analysis applied to the 400\us\ ``test'' exposure 
gives the qunybble state as \{1110\} with estimated probability of error $\approx \sum_{k=1}^4 \mbox{e}^{-R_k} = 1.4\times 10^{-11}$. (b) Readout 
error per qubit averaged over trials of all 16 qunybble states $\{0000,\ldots,1111\}$, whose distribution is shown in the inset. The error due to 
\lev{D}{5/2} decay is excluded as described in the text. Analysis methods: (T)hreshold; (M)aximum likelihood ignoring neighbours; (MN) iterative  
maximum likelihood method taking into account nearest neighbours, with (solid line) a Monte-Carlo simulation of $10^9$ trials. Adjacent ROIs overlap 
when their diameters exceed the 14\um\ mean ion spacing.
}
\label{F:nyberror}
\end{figure}

A single trial consists of 6 identical camera exposures each of 400\us\ duration (fig.~\ref{F:nyberror}a), with the CCD being read out between each. 
An initial ``check'' exposure is taken with all ions fluorescing. Then a short (12\us) pulse of 393\nm\ light shelves the ions in \lev{D}{5/2} with a 
probability $0.46$ per ion, to prepare a random qunybble state (with each state approximately equally probable, inset fig.~\ref{F:nyberror}b). The 
next two exposures are added to give the ``pre'' measurement, the fourth exposure constitutes the actual readout ``test'' measurement, and the last 
two exposures are added for the ``post'' measurement. Finally an 854\nm\ laser pulse returns all ions to the bright state. 

This sequence was repeated 30100 times, giving 120400 qubit trials. We first identify those trials where the ``pre'' and ``post'' measurements 
disagree, and discard them, using thresholding of pixel counts summed over rectangular ROIs. Two thresholds were used to improve our confidence in the 
retained trials: we demand either that the ``pre'' and ``post'' measurements are {\em both\/} above the upper threshold, or that {\em both\/} are 
below the lower threshold. The separation between the upper and lower thresholds was such that 95\% of trials were retained~\footnote{A single 
threshold keeps 99\% of trials, and gives $\epsilon_X = 0.3(2)\times 10^{-4}$.}.

Having thus selected a set of trials with known qubit states, we proceed to analyse the ``test'' exposures. We first apply a simple threshold 
analysis, treating each ion independently. The pixel counts $n_{ki}$ for each ion $k$ are summed over independent brightness-order ROIs, and an 
independent threshold is chosen for each ion. The minimum readout error (per qubit) is $\epsilon_X=6.8(8)\times 10^{-4}$ (fig.~\ref{F:nyberror}b) and 
$\epsilon_X$ rises rapidly above $N\approx 25$ pixels, where the ROI diameters exceed the mean ion spacing.

The spatial maximum likelihood analysis should be more immune to the effects of image cross-talk since it requires fewer pixels and gives less weight 
to pixels further from the ion of interest. For each ion $k$ we compare the likelihoods $p_B=\prod_{i=1}^N B_{ki}(n_{ki})$ and $p_D=\prod_{i=1}^N 
D_{ki}(n_{ki})$ that the set of counts $\{n_{ki}\}$ came from a bright or dark ion, where $B_{ki}(n_{ki})$ is the probability of observing $n_{ki}$ 
counts on ion $k$'s $i$th brightest pixel when that ion is bright, and similarly for $D_{ki}(n_{ki})$. The probability distributions $B_{ki}$ and 
$D_{ki}$ are obtained from the ``pre'' and ``post'' exposures only and are thus independent of the ``test'' exposures. Results are shown in 
fig.~\ref{F:nyberror}b and reach an error $\epsilon_X=0.7(3)\times 10^{-4}$ per qubit, an order of magnitude below simple thresholding. As expected, 
the error is less sensitive to the ROI overlapping neighbouring ions.

Nevertheless, we can improve the analysis by taking into account the effect of neighbouring ions explicitly. There is a significant effect on the dark 
distribution $D_k$ depending on whether the neighbour $k'$ is bright or dark; indeed for a pixel far from ion $k$ but close to ion $k'$ the shape of 
the distribution will be dominated by the state of $k'$ even though the pixel may still contain useful information on the state of $k$. To extract 
this information, we now allow independent probability distributions $B_{\nu ki}(n_{ki})$ and $D_{\nu ki}(n_{ki})$ for each neighbour state 
$\{\nu\}=\{00,01,10,11\}$. These distributions are again found using the ``pre'' and ``post'' exposures only, where we know the state of both ion $k$ 
and its neighbours $\{\nu\}$. However, when we use these distributions to analyse a particular ``test'' exposure we do {\em not\/} assume any 
knowledge of the neighbour states: instead we proceed iteratively, making an initial guess for the qunybble state of $\{0000\}$, measuring the state 
of each qubit $k$ with $\{\nu\}=\{00\}$, recalculating using the neighbour states from the first iteration to give a second measurement, and so on 
until the inferred qunybble state is stable. Results from this iterative maximum likelihood analysis are also plotted in fig.~\ref{F:nyberror}b: for 
ROI size $N\ge 60$ pixels the readout error per qubit reaches an average level of $\epsilon_X=0.1(1)\times 10^{-4}$ and is even zero (in 114606 
trials) for certain $N$. There is now no penalty for allowing the ROI to overlap neighbouring ions. The agreement with the simulation in 
fig.~\ref{F:nyberror}b indicates that systematic effects (e.g., correlated errors common to ``pre'', ``post'' and ``test'' exposures) are below the 
statistical uncertainty. 

Finally we can search for effects due to non-nearest neighbours by generalizing the preceding analysis to neighbour states 
$\{\nu\}=\{000,\ldots,111\}$, which allows for all permutations of the three neighbour ions to any ion $k$. We find no reduction in $\epsilon_X$ or 
$N$ and conclude that any effect is at or below the $\ish 10^{-5}$ level.

The single-ion experiment achieved a readout fidelity of 99.991(3)\% with a single 400\us\ camera exposure, limited by the photon collection 
efficiency and the lifetime of the \lev{D}{5/2} qubit state. The four-ion qunybble experiment demonstrated that, when the effects of the \lev{D}{5/2} 
decay are excluded, discrimination accuracy of 99.999(1)\% can be achieved with the same exposure time, even in the presence of significant image 
cross-talk from neighbouring ions. Together these two experiments imply that 99.99\% net fidelity can be attained with the camera for simultaneous 
multiple qubit readout. Since it was found sufficient to include only the effects of nearest neighbors, this result should extend to one-dimensional 
qubit registers of arbitrary length, given comparable imaging quality over the entire ion string. As the error due to neighbour effects is small 
($\epsilon_X\ll\epsilon$), the fidelity should be similar even for a 2D array of ions (or neutral atoms in a long-wavelength optical 
lattice~\cite{07:Nelson}, given the same fluorescence and dark$\leftrightarrow$bright rates). This $512\times 512$ pixel camera could be used to 
measure a 2D array of at least 10000 qubits simultaneously; the exposure time would then be negligible compared with the CCD readout time (\ish 1\us\ 
per qubit at the fastest available readout speed of 0.1\us/pixel). A CCD which allowed parallel pixel readout would offer a significant speed-up. 
Image processing time is $<0.1$\us/pixel, so the main classical computing resource cost is storage (\ish 1\unit{kB}/pixel) for the $B_i$ and $D_i$ 
distributions.

The cross-talk in our system was limited by imperfections in the imaging system. To estimate what is theoretically possible, we plot on 
fig.~\ref{F:Xtalk}b the calculated image cross-talk for a diffraction-limited imaging system at $\lambda=397\nm$ with the same numerical aperture 
$\sin\alpha=0.25$, but with ion spacings reduced by a factor of 10. The cross-talk from neighbouring ions was calculated by integrating the Airy 
pattern over off-axis circular ROIs, and is qualitatively similar to our measured cross-talk. In this calculation the spacing between the ions, 
1.4\um, is less than the diameter of their Airy discs $\approx 1.22\lambda/\tan\alpha=1.9\um$. The implication is that readout fidelity at the 99.99\% 
level is attainable for ions or atoms spaced close to the diffraction limit, with an imaging system possessing the same collection efficiency.

We thank Andor Technology for the loan of the camera (model DU-897), and other members of the Oxford ion trap group (particularly D.~N.~Stacey and 
A.~M.~Steane) for helpful discussions. This work was funded by EPSRC (QIP IRC), IARPA (ref.\ 47968-PH-QC), the European Commission (SCALA, MicroTrap) 
and the Royal Society. 

Related experiments have recently been performed by F.~Z\"{a}hringer and co-workers at Innsbruck University.

\end{document}